\documentstyle[twoside,fleqn,espcrc2]{article}
\title{
Structure of flux tube in $SU(2)$ lattice gauge theory
}
\author{
Hiroshi Shiba 
\address{
Department of Physics, Kanazawa University, Kanazawa 920-11, Japan
}}

\begin{document}

\begin{abstract}
The structure of the flux tube is studied in $SU(2)$ QCD 
from the standpoint of the abelian projection theory. 
It is shown that the flux distributions of 
the orthogonal electric field and the magnetic field 
are produced by the effect that 
the abelian monopoles in the maximally abelian (MA) gauge 
are expelled from the string region. 
\end{abstract}

\maketitle

\input epsf

\section{Introduction}
It is important 
to investigate the structure of the flux tube 
in order to understand the confinement mechanism. 
A possible mechanism is 
the dual Meissner effect due to the condensation 
of the abelian monopoles, which is the basic idea of 
the abelian projection theory\cite{thooft2}. 
From the standpoint of this theory, 
the flux distribution in the flux tube 
is explained by the following two effects: 

1. The effect that the abelian monopoles are 
expelled from the string region. 

2. The squeezing of the abelian electric flux 
into the string region. 

The purpose of this work is to check 
the validity of this picture (in the MA gauge). 

\section{Color distribution around a monopole}
If monopoles have no correlation with color fields, 
they can not contribute directly to the color distribution 
in the flux tube. 
In order to investigate the color distribution 
around a monopole, we introduce a correlation 
function 
%\begin{eqnarray}
%\vspace{-2.5pt}
$$
p^{mon}_{\mu\nu}({\bf r})=\frac{
\langle \delta^{mon}(c({\bf x},x_0))
P_{\mu\nu}({\bf x} +{\bf r},x_0)\rangle
}
{\langle \delta^{mon} \rangle}
-\langle P_{\mu\nu} \rangle, 
%\end{eqnarray}
$$
%\vspace{-2.5pt}
with 
$
P_{\mu\nu}(x) = 1 - \frac{1}{2} tr
[U_{\mu}(x) U_{\nu}(x+\hat{\mu})
 U^{\dagger}_{\mu}(x+\hat{\nu}) U^{\dagger}_{\nu}(x)]$ 
being the plaquette action. 
$c({\bf x},x_0)$ is a spatial cube located 
at $({\bf x},x_0)$ and $\delta^{mon}(c)$ is an operator defined by 
%\vspace{-2.5pt}
$$
\hspace{-8.5mm}
%\begin{eqnarray}
\delta^{mon}(c)= \left\{ \begin{array}{ll}
1 & \mbox{if $c$ contains a monopole,}\\ 
0 & \mbox{otherwise.}
\end{array}
\right.
%\end{eqnarray}
$$
%\vspace{-3pt}
$p^{mon}_{\mu\nu}({\bf r})$ 
measures the difference between the average 
of the plaquette action in the presence of a monopole, 
separated by ${\bf r}$, and its vacuum expectation. 
We performed Monte Carlo simulations on a $16^4$ lattice 
at $\beta = 2.5$. 
The Degrand-Toussaint scheme was used for determining 
the magnetic charge in a cube from abelian gauge fields 
after the abelian projection in the MA gauge\cite{suzu93}. 

\begin{figure}[htb]
%\vspace{3pt}
%\framebox[55mm]{\rule[-21mm]{0mm}{43mm}}
\epsfxsize=0.5\textwidth
\vspace{-60pt}
\begin{center}
\leavevmode
\epsfbox{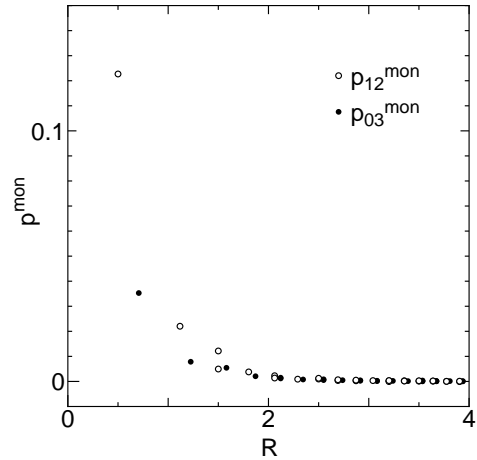}
\end{center}
\vspace{-55pt}
\caption{
The correlation $p^{mon}_{\mu\nu}$ versus $R$. 
}
%\label{fig:fig1}
\label{fig1}
\vspace{-30pt}
\end{figure}

In Fig.\ref{fig1}, 
%In Fig.1, 
$p^{mon}_{\mu\nu}$ is plotted against $R$, the distance 
from the center of the cube to the plaquette. 
Monopoles correlate to color fields around themselves: 
the average of the plaquette action is large in the 
vicinity of a monopole and decreases to the 
vacuum expectation as the distance becomes large 
$(R > 2)$. 
%---------- fig1 ---------
%-------------------------

\section{Monopole contribution to the flux tube}
The flux distribution in the presence of a static 
$q \bar{q}$ pair is measured by the correlation function: 
$$
\hspace{-9.6mm}
%\begin{eqnarray}
p_{\mu\nu}({\bf x})=\frac{
\langle W(R,T)\ P_{\mu\nu}({\bf x},T/2)\rangle
}
{\langle W(R,T) \rangle}
- \langle P_{\mu\nu} \rangle,
%\end{eqnarray}
$$
where $W(R,T)$ is a Wilson loop of the size $R \times T$. 
Position of the plaquette 
relative to the $q \bar{q}$ sources is specified by ${\bf x}$. 
The squared electric and magnetic field components 
(in Minkowski space) relative to the vacuum 
are given by 
%\vspace{-3.5pt}
$$
\hspace{-22mm}
%\begin{eqnarray}
(E^{2}_{1},E^{2}_{2},E^{2}_{3})
=-2 \frac{\beta}{a^{4}}(p_{01},p_{02},p_{03}),
$$
%\vspace{-3.5pt}
$$
\hspace{-24.6mm}
(B^{2}_{1},B^{2}_{2},B^{2}_{3})
= 2 \frac{\beta}{a^{4}}(p_{23},p_{31},p_{12}).
%\end{eqnarray}
$$
%\vspace{-3.5pt}
From these fields, the energy and action densities 
are obtained: 
%\vspace{-3.5pt}
$$
\hspace{-22.2mm}
%\begin{eqnarray}
{\cal E}=\frac{1}{2}[{\bf E}^{2}+{\bf B}^{2}],\ 
{\cal S}=\frac{1}{2}[{\bf E}^{2}-{\bf B}^{2}].
%\end{eqnarray}
$$
%\vspace{-3.5pt}
An interesting feature of the flux distribution measured 
is that the magnetic energy density relative to the 
vacuum is negative in contrast to the positive 
electric energy density. 
% A dual analogy to the Nielsen--Olesen string 
% was suggested by Sommer\cite{sommer}. 
From the standpoint of the abelian projection theory, 
it is expected that this feature can be explained 
by abelian monopoles. 
The monopoles, which possibly correspond to Cooper Pairs 
in a superconductor, may be expelled from the string 
region of the flux tube. 
This indicates that 
the monopole density is lower in the inner 
part of the flux tube than in the vacuum. 
Such a behavior is also observed on the lattice, 
in the MA gauge\cite{debbio}. 
In addition, the average of the plaquette action 
is large in the vicinity of a monopole, 
as shown in the previous section. 
As a result, the monopoles can make negative contributions 
to the correlation $p_{\mu\nu}({\bf x})$ in the string region. 
This means that the monopole contributions are positive 
for the electric energy density and are negative for 
the magnetic energy density.

%========   six figures start ! =============
\begin{figure*}[tbh]

\vspace{-30pt}

%---------- three figures start ! ---------
\begin{center}

\epsfxsize=0.35\textwidth
\leavevmode
\epsfbox{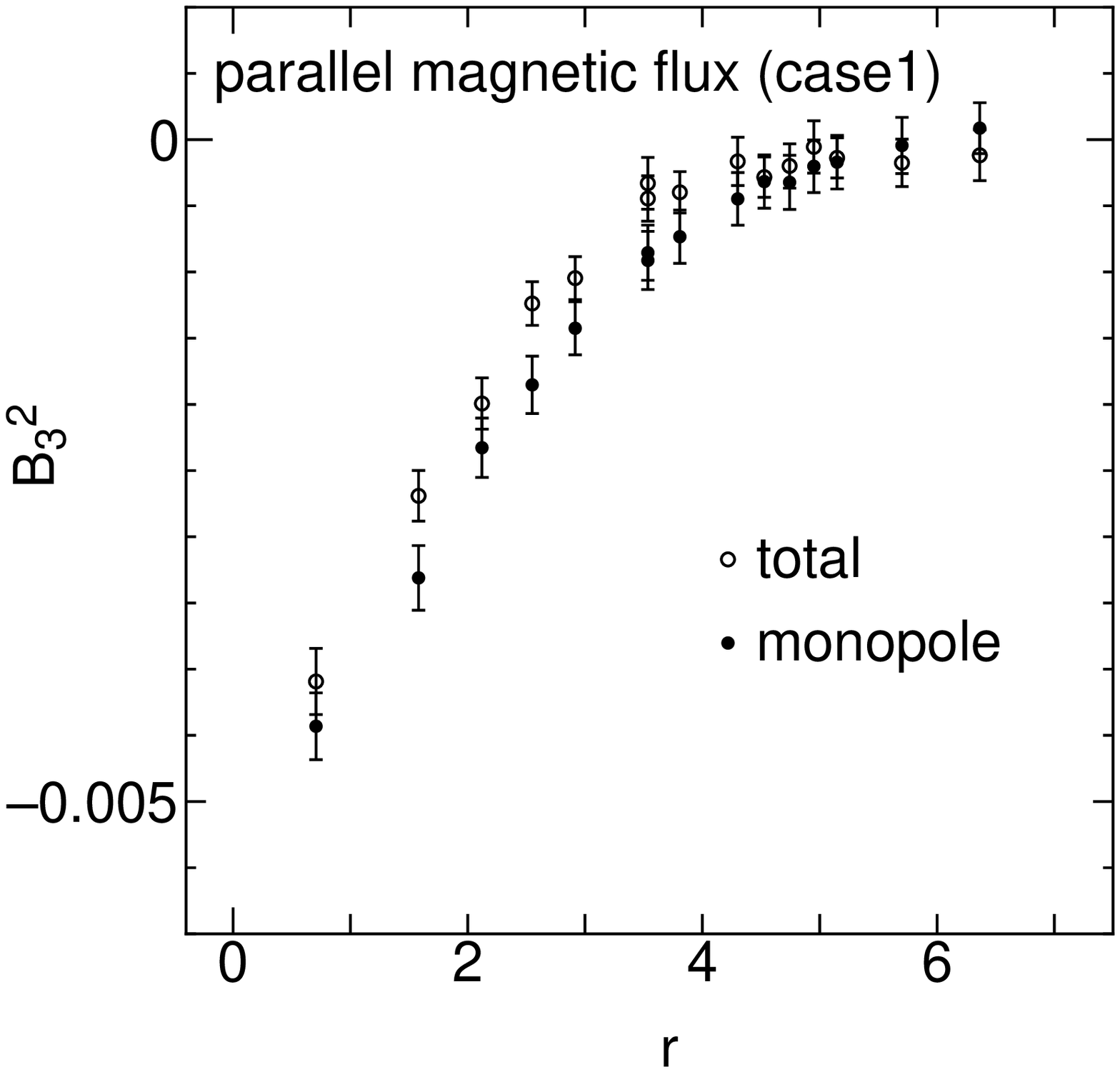}

\epsfxsize=0.35\textwidth
\leavevmode
\epsfbox[520 -467 1115 129]{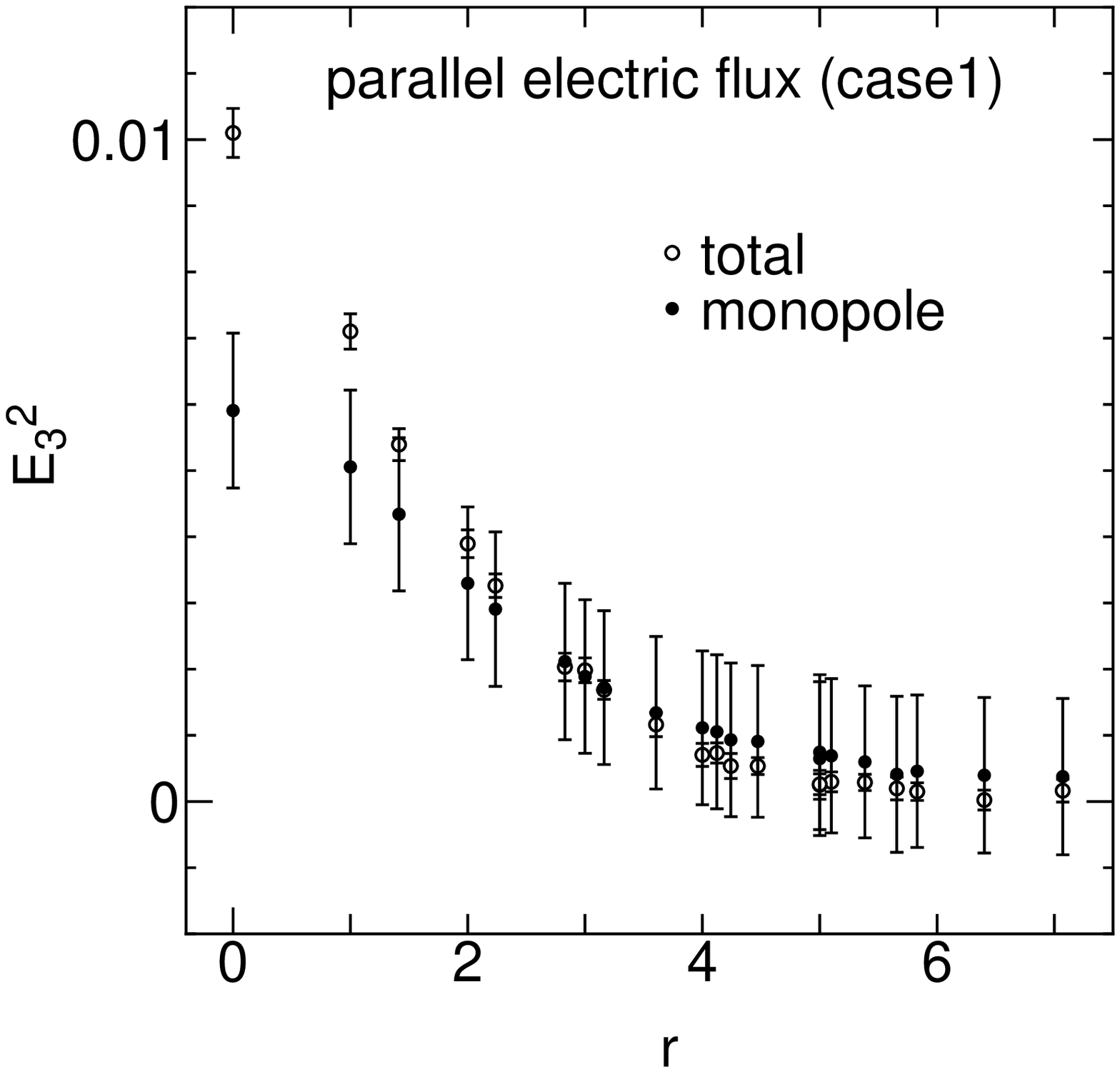}

\epsfxsize=0.35\textwidth
\leavevmode
\epsfbox[-520 -1009 75 -471]{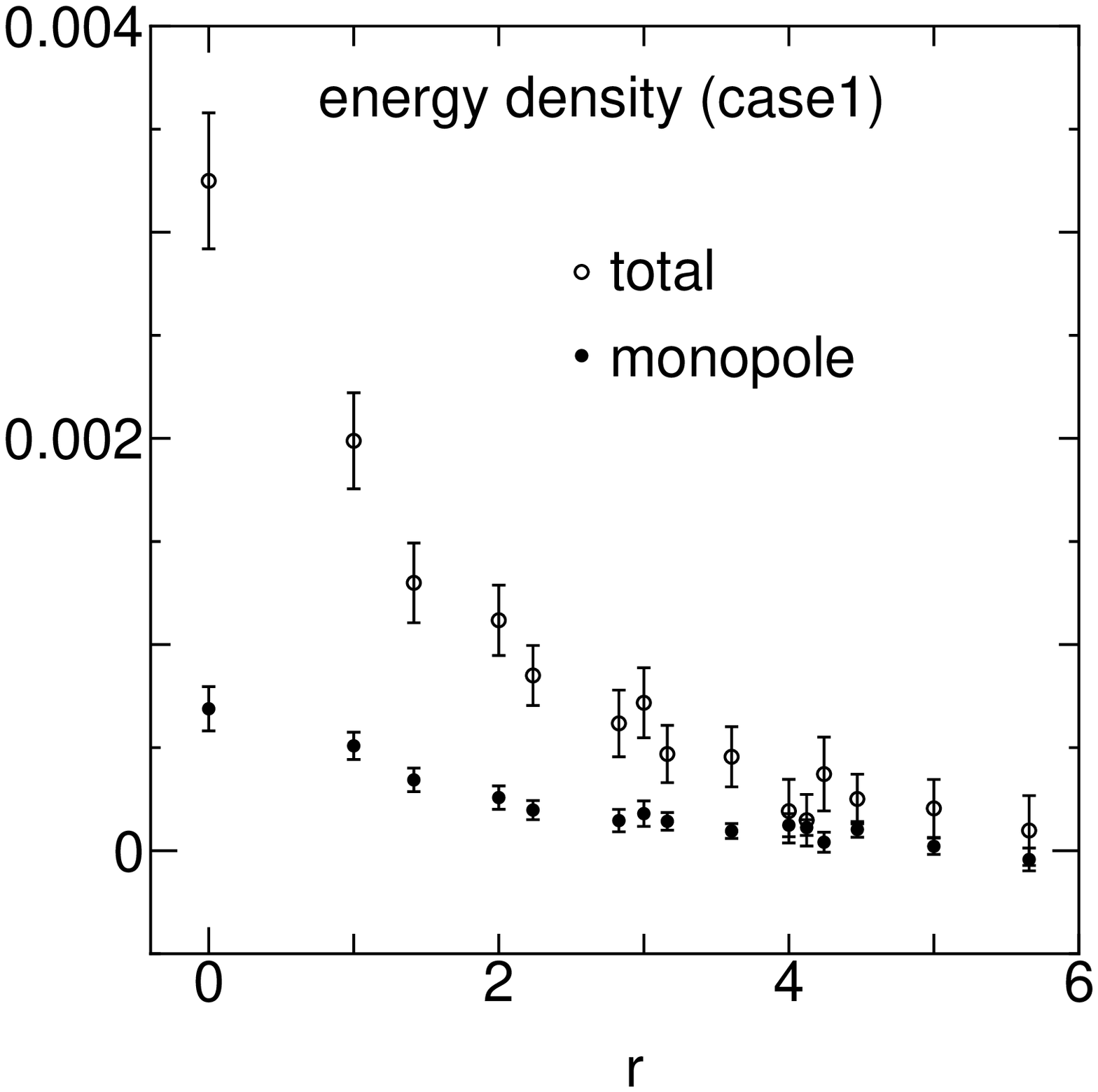}

\end{center}

\vspace{-360pt}

%---------- three figures end ---------
%---------- three figures start ! ---------
\begin{center}

\epsfxsize=0.35\textwidth
\leavevmode
\epsfbox{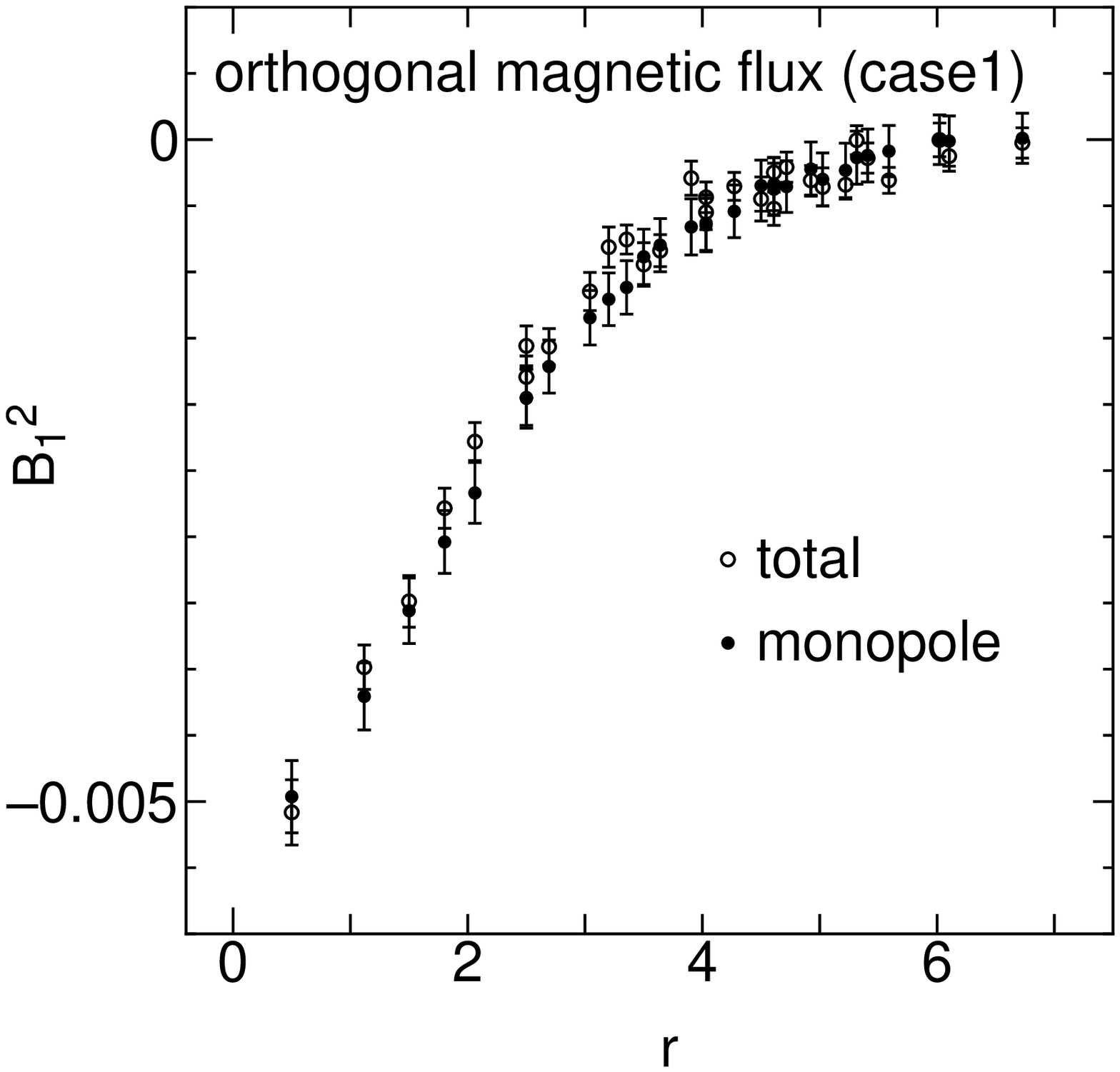}

\epsfxsize=0.35\textwidth
\leavevmode
\epsfbox[520 -467 1115 129]{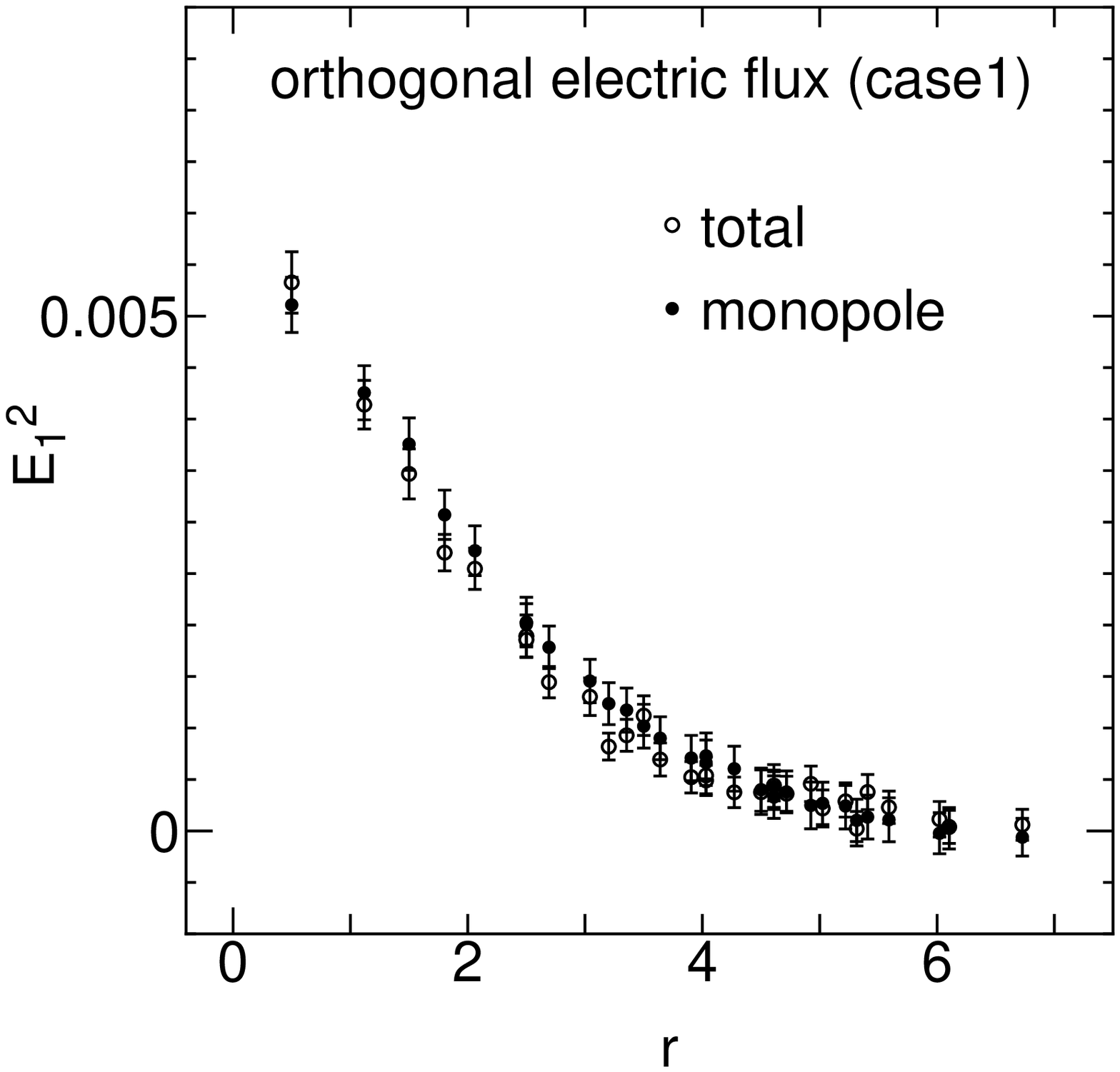}

\epsfxsize=0.35\textwidth
\leavevmode
\epsfbox[-520 -1009 75 -471]{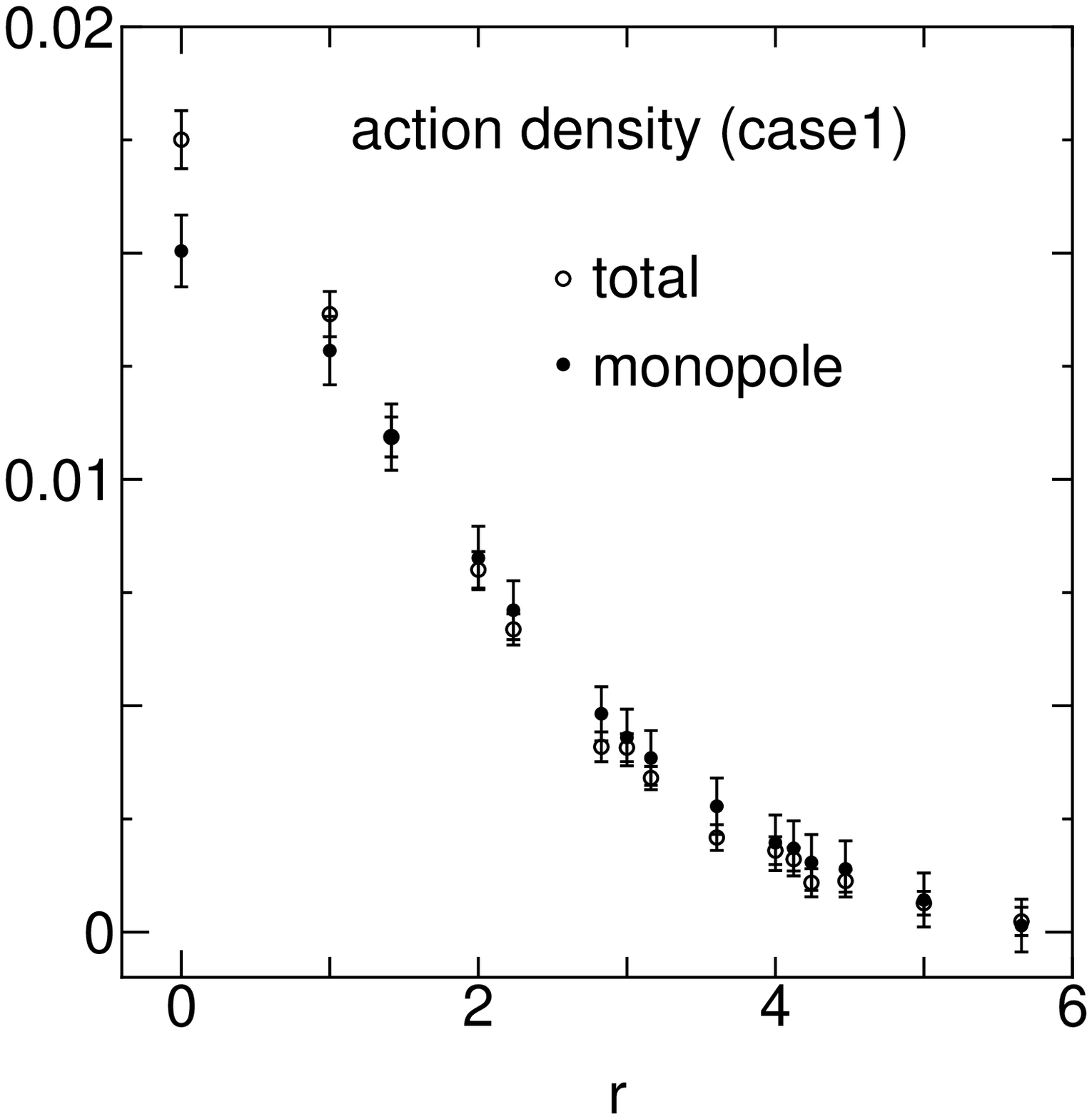}

\end{center}

\vspace{-350pt}

%---------- three figures end ---------

\caption{
Flux distributions and monopole contributions. 
$r$ is the transverse distance 
from the $q\bar{q}$ axis.
}
\label{fig:histogram}

\vspace{-10pt}

\end{figure*}
%========   six figures end =============

To determine the monopole contributions quantitatively, 
we separate the whole lattice into the core regions 
corresponding to the monopole cores 
and the interstitial regions between them 
on the basis of the observation in the previous section. 
We define operators 
$\delta^{I}_{\mu\nu}(x)$ and $\delta^{I \hspace{-1mm} I}_{\mu\nu}(x)$ 
on a plaquette by 
%\vspace{-3pt}
$$
%\begin{eqnarray}
\delta^{I}_{\mu \nu}(x) = \left\{ \begin{array}{ll}
         1 & \mbox{if plaquette $(x;\mu,\nu)$ belongs to} \\
           & \mbox{the core regions,} \\
         0 & \mbox{otherwise,} \end{array} \right. 
%\end{eqnarray}
$$
\vspace{-3pt}
%and 
$$
\hspace{-44.4mm}
%\begin{eqnarray}
\delta^{I \hspace{-1mm} I}_{\mu \nu}(x) = 1-\delta^{I}_{\mu \nu}(x).
%\end{eqnarray}
$$
%\vspace{-3pt}
The correlation $p_{\mu\nu}({\bf x})$ can be decomposed 
into $p_{\mu\nu}({\bf x}) = 
p^{M}_{\mu\nu}({\bf x})+
p^{F}_{\mu\nu}({\bf x})$, where 
$$
\hspace{-12mm}
%\begin{eqnarray}
p^{M}_{\mu \nu}({\bf x}) = 
\sum_{i}[ \langle \delta^{i}_{\mu \nu}({\bf x}) \rangle_{q \bar{q}}
- \langle \delta^{i}_{\mu \nu} \rangle ] \langle P^{i}_{\mu \nu}
\rangle_{vac} ,
$$
$$
\hspace{-3.7mm}
p^{F}_{\mu \nu}({\bf x}) = 
\sum_{i} \langle \delta^{i}_{\mu \nu}({\bf x}) \rangle_{q \bar{q}}
[\langle P^{i}_{\mu \nu}({\bf x}) \rangle_{q \bar{q}} -
\langle P^{i}_{\mu \nu} \rangle_{vac}] ,
$$
$$
\hspace{-20mm}
\langle \delta^{i}_{\mu \nu}({\bf x}) \rangle_{q \bar{q}} =
\frac{ \langle W(R,T) \delta^{i}_{\mu \nu}({\bf x}, T/2) \rangle}
{ \langle W(R,T) \rangle } ,
$$
$$
\hspace{-33.2mm}
\langle P^{i}_{\mu \nu} \rangle_{vac} =
\frac{ \langle \delta^{i}_{\mu \nu}(x) P_{\mu \nu}(x) \rangle}
{ \langle \delta^{i}_{\mu \nu} \rangle } ,
$$
$$
\hspace{-1.2mm}
\langle P^{i}_{\mu \nu}({\bf x}) \rangle_{q \bar{q}} =
\frac{ \langle W(R,T) \delta^{i}_{\mu \nu}({\bf x},T/2) 
P_{\mu \nu}({\bf x},T/2) \rangle}
{ \langle W(R,T) \delta^{i}_{\mu \nu}({\bf x},T/2) \rangle } ,
%\end{eqnarray}
$$
with $i = I, I \hspace{-1mm} I$. 
$p^{M}_{\mu\nu}({\bf x})$ monitors how the change in the monopole 
distribution by the presence of a $q \bar{q}$ pair 
$\langle \delta^{i}_{\mu\nu}({\bf x}) \rangle_{q \bar{q}} -
 \langle \delta^{i}_{\mu\nu} \rangle$ 
contributes to $p_{\mu\nu}({\bf x})$. 
It corresponds to the monopole contributions. 
On the other hand, $p^{F}_{\mu\nu}({\bf x})$ monitors how the difference 
$\langle P^{i}_{\mu\nu}({\bf x}) \rangle_{q \bar{q}} -
 \langle P^{i}_{\mu\nu} \rangle_{vac}$, 
which is irrelevant to the 
monopole distribution, contributes to $p_{\mu\nu}({\bf x})$. 
From the standpoint of the abelian projection theory, 
this contribution corresponds 
to the (abelian) net electric flux, which spreads 
from $q \bar{q}$ sources and connects them. 
We call $p^{F}_{\mu\nu}({\bf x})$ as the net flux contributions. 
If our expectation is true, $p^{F}_{\mu\nu}({\bf x})$ must 
contribute only to the electric field component 
parallel to the $q \bar{q}$ axis in the flux tube; 
in other words, the electric field components 
perpendicular to the $q \bar{q}$ axis and the 
magnetic field components should 
be reproduced by the monopole contributions 
alone. 

We have measured the total flux and 
the monopole contributions on a $16^4$ lattice at $\beta = 2.5$. 
284 configurations were taken for the measurements. 
The flux distribution on the transverse plane 
midway between the $q \bar{q}$ pair was measured 
on the Wilson loop of the size $R \times T$ with $R = 6$, 
$T = 5$. 
Smearing was applied to the spatial part of the 
Wilson loop in order to increase the ground 
state overlap\cite{bali}. 
After 10 smoothing steps we reached 
the overlap of $78 \%$ for a spatial separation $R = 6$. 
All plaquettes were separated into the core regions 
and the interstitial regions. 
To check the dependence on the way of this 
separation, we tried three types of separation. 
In case 1, the core regions consists of 
plaquettes on the boundary of a cube 
that contains monopoles. 
In case 2 and in case 3, additional plaquettes 
were included to the core regions. 
But quantitative results did not depend 
on this choice.

%In Fig.\ref{fig2}, the total flux and the 
In Fig.2, the total flux and the 
monopole contributions are compared. 
The monopole contributions almost agree with 
the total flux for the orthogonal electric 
flux $E^{2}_{1}$ and the magnetic flux $B^{2}_{3}$ and $B^{2}_{1}$. 
As is predicted by the abelian projection theory, 
the flux distribution of these field components 
is produced by the effect that the monopoles 
are expelled from the string region of the 
flux tube. 
The monopoles also contribute to the parallel 
electric flux $E^{2}_{3}$. 
The monopole contributions to $E^{2}_{3}$ are 
approximately equal 
to the contributions to the other flux components 
in magnitude. 
As for the total flux, 
the parallel electric flux is larger than the 
other flux components in magnitude. 
As a result, the monopole contributions 
are less than the total for the parallel electric flux 
by the amount of this difference. 
The difference between them comes from the net 
flux contributions. 
As for the energy density, 
the monopole contributions are far less than the total. 
The substantial difference comes from 
that in the parallel electric flux. 
This result shows that the energy is 
almost carried by the net flux contributions. 
On the other hand, 
the monopole contributions are dominant 
for the action density. 

Our results support the abelian 
projection theory in the MA gauge. 
To prove the abelian dominance in the 
flux tube, it is necessary to show 
that the net flux contributions to the parallel 
electric flux are composed of the abelian 
flux alone. 
Such an investigation is in progress.


\vspace{-5pt}

\begin{thebibliography}{99}
\bibitem{thooft2} G. 'tHooft, Nucl. Phys. {\bf B190}, 455 (1981).
\bibitem{suzu93} T. Suzuki, Nucl. Phys. 
B(Proc. Suppl.) {\bf 30}, 176 (1993) 
and references therein.  
% \bibitem{sommer} R.Sommer, Nucl. Phys. {\bf B291}, 673 (1987).
\bibitem{debbio} L.D.Debbio $et$ $al.$, Phys. Lett. 
{\bf 267B}, 254 (1991).
\bibitem{bali} G.S.Bali $et$ $al.$, 
Nucl. Phys. B(Proc. Suppl.) {\bf 34}, (1994) 216 .
\end{thebibliography}
\end{document}